\documentclass[acmtog]{acmart}
\acmSubmissionID{478}

\usepackage{booktabs} 

\usepackage[ruled]{algorithm2e} 

\SetAlFnt{\small}
\SetAlCapFnt{\small}
\SetAlCapNameFnt{\small}
\SetAlCapHSkip{0pt}

\copyrightyear{2025}
\acmYear{2025}
\setcopyright{rightsretained}
\acmConference[XXX]{XXX}{XXX}{XXX}
\acmBooktitle{XXX}
\acmDOI{XXX}
\acmISBN{XXX}

\usepackage{graphicx}
\usepackage{amsmath}
\usepackage{mathtools}
\usepackage{caption}
\usepackage{float}
\usepackage{enumitem}
\usepackage{bbm}
\usepackage{wrapfig}
\usepackage{subfigure}
\usepackage{multirow}
\usepackage{comment}

\usepackage[utf8]{inputenc} 
\usepackage[T1]{fontenc}    
\usepackage{hyperref}       
\usepackage{url}            
\usepackage{booktabs}       
\usepackage{amsfonts}       
\usepackage{nicefrac}       
\usepackage{microtype}      
\usepackage{xcolor}         

\usepackage{cleveref}

\Crefname{equation}{Eq.}{Eqs.}
\Crefname{figure}{Fig.}{Figs.}
\Crefname{table}{Tab.}{Tabs.}
\Crefname{section}{Sec.}{Secs.}

\newcommand{\parag}[1]{\paragraph{#1}}
\newcommand{\Teta}[0]{\mathbf{\Theta}}

\newcommand{\bx}{\mathbf{x}}
\newcommand{\bz}{\mathbf{z}}
\newcommand{\bu}{\mathbf{u}}
\newcommand{\bX}{\mathbf{X}}

\newcommand{\bV}{\mathbf{V}}
\newcommand{\bS}{\mathbf{S}}
\newcommand{\bD}{\mathbf{D}}
\newcommand{\bN}{\mathbf{N}}
\newcommand{\bC}{\mathbf{C}}
\newcommand{\bm}{\mathbf{m}}
\newcommand{\bG}{\mathbf{G}}

\begin{document}
\title{Single View Garment Reconstruction Using Diffusion Mapping Via Pattern Coordinates}

\author{Ren Li}
\authornote{Corresponding author is Ren Li (ren.li@epfl.ch).}
\orcid{0000-0003-2998-7104}
\affiliation{%
  \institution{École Polytechnique Fédérale de Lausanne}
  \country{Switzerland}}
\email{ren.li@epfl.ch}
\author{Cong Cao}
\orcid{0009-0001-5989-8367}
\affiliation{%
  \institution{Mohamed bin Zayed University of Artificial Intelligence}
  \country{United Arab Emirates}}
\email{cong.cao@mbzuai.ac.ae}
\author{Corentin Dumery}
\orcid{0000-0001-5314-7979}
\affiliation{%
 \institution{École Polytechnique Fédérale de Lausanne}
 \country{Switzerland}}
\email{corentin.dumery@epfl.ch}
\author{Yingxuan You}
\orcid{0000-0002-0154-4590}
\affiliation{%
 \institution{École Polytechnique Fédérale de Lausanne}
 \country{Switzerland}}
\email{yingxuan.you@epfl.ch}
\author{Hao Li}
\orcid{0000-0002-4019-3420}
\affiliation{%
  \institution{Pinscreen, Mohamed bin Zayed University of Artificial Intelligence}
  \country{United Arab Emirates}}
\email{hao@hao-li.com}
\author{Pascal Fua}
\orcid{0000-0002-6702-9970}
\affiliation{%
 \institution{École Polytechnique Fédérale de Lausanne}
 \country{Switzerland}}
\email{pascal.fua@epfl.ch}

\ccsdesc[500]{Computing methodologies}
\ccsdesc[500]{Computing methodologies~Reconstruction}
\ccsdesc[500]{Computing methodologies~Shape inference}

\keywords{3D Garment Reconstruction, Image based Modeling, Diffusion Models}
\begin{teaserfigure}
    \includegraphics[width=\textwidth]{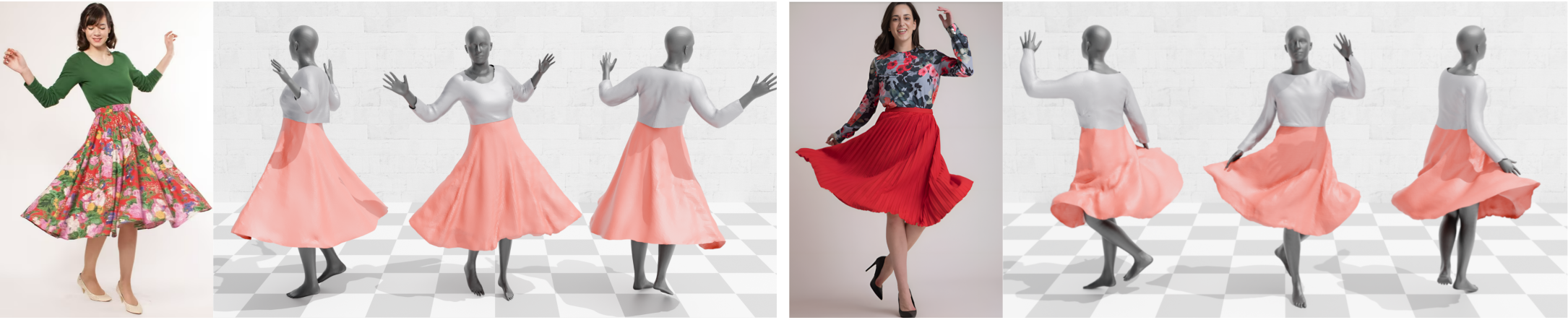}
    \caption{Given a single image of a clothed person, our proposed method can reconstruct high-fidelity 3D garment models with realistic details.}
    \label{fig:teaser}
\end{teaserfigure}


\begin{abstract}

Reconstructing 3D clothed humans from images is fundamental to applications like virtual try-on, avatar creation, and mixed reality. While recent advances have enhanced human body recovery, accurate reconstruction of garment geometry—especially for loose-fitting clothing—remains an open challenge. 
We present a novel method for high-fidelity 3D garment reconstruction from single images that bridges 2D and 3D representations. Our approach combines Implicit Sewing Patterns (ISP) with a generative diffusion model to learn rich garment shape priors in a 2D UV space. A key innovation is our mapping model that establishes correspondences between 2D image pixels, UV pattern coordinates, and 3D geometry, enabling joint optimization of both 3D garment meshes and the corresponding 2D patterns  by aligning learned priors with image observations.
Despite training exclusively on synthetically simulated cloth data, our method generalizes effectively to real-world images, outperforming existing approaches on both tight- and loose-fitting garments. The reconstructed garments maintain physical plausibility while capturing fine geometric details, enabling downstream applications including garment retargeting and texture manipulation. 

\end{abstract}

\maketitle

\section{Introduction}
\label{sec:intro}

Recovering the pose and shape of people's bodies, along with the shape of their garments, solely from images has many applications. They include fashion design, virtual try-on, creating 3D avatars, telepresence, and immersive VR/AR. 
Recent years have seen tremendous progress in modeling people wearing tight-fitting clothing both in terms of body poses~\cite{Bogo16,Lassner17a,Kanazawa18a,Kolotouros19,Pavlakos19a,Georgakis20,Moon20,Li21i,Li21g,xu2024,li2024hyre} and 3D shape of the clothes~\cite{Danerek17,Bhatnagar19,Jiang20d,Corona21,Moon22,Li22c,DeLuigi23,Li23a}. However, accurately modeling clothing, especially when it fits loosely on the body, remains a challenge. Most current work relies on a single 3D model to jointly represent the body and its clothing. While this  can produce visually impressive reconstruction results, this fused representation of humans and their garments makes it impossible to perform realistic cloth simulation or virtual try-on. 


\begin{figure*}[ht!]
    \centering
    \includegraphics[width=.98\textwidth]{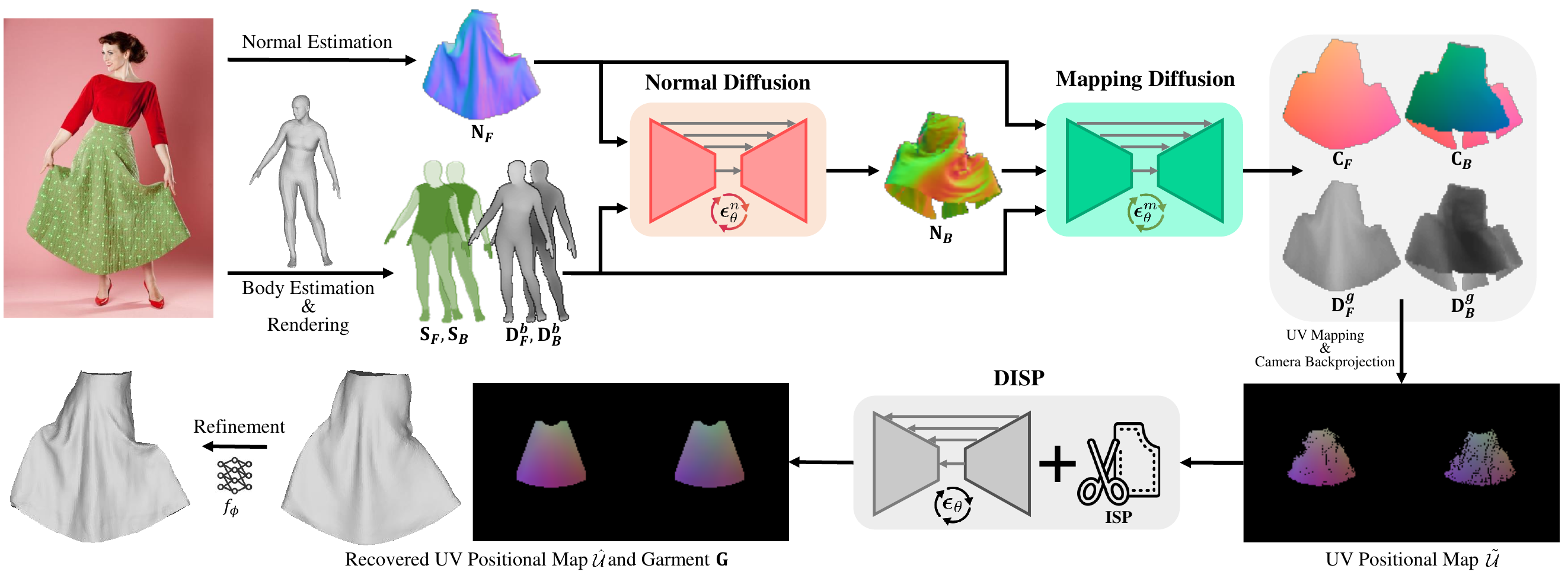}
    \caption{\textbf{Pipeline}.  Given an image of a clothed person, we first estimate the front normal $\bN_F$ of the target garment, and the SMPL body model which is used to render the body part segmentation ($\bS_F$, $\bS_B$) and depth ($\bD_F^b$, $\bD_B^b$) images. The back normal $\bN_B$ of the garment is estimated subsequently by the diffusion model $\boldsymbol{\epsilon}_{\theta}^n$. We then predict the UV-coordinate ($\bC_F$, $\bC_B$) and the depth ($\bD_F^g$, $\bD_B^g$) images from the garment normal and body estimations with the mapping model $\boldsymbol{\epsilon}_{\theta}^m$. The incomplete UV positional map $\Tilde{\mathcal{U}}$ is produced from them using the camera backprojection. Finally, we fit $\Tilde{\mathcal{U}}$ to DISP to recover the complete UV positional map $\hat{\mathcal{U}}$ and the corresponding garment mesh $\bG$, which is further improved by the refinement.}
    \label{fig:pipe}
\end{figure*}

Thus, independent body and garment models are needed. Such modeling is made difficult by the intricate structure of clothing. Since garments are thin surfaces with near-infinite degrees of freedom, they undergo complex deformations caused by dynamic factors. The many design styles and shape variations of clothing items introduce further complexity, making the modeling process even more challenging and the acquisition of real 3D data more difficult. In turn, this impedes the deployment of learning-based methods for garment reconstruction. To address these challenges, several works~\cite{Danerek17,Bhatnagar19,Jiang20d,Casado22,Liu23b} rely on pre-designed mesh templates to define the garment geometry, and employ linear blend skinning (LBS)~\cite{Loper15} of the underlying body model to capture deformations caused by body motion. However, this requires mesh templates for the clothes, which limits modeling flexibility and generality. Furthermore, while skinning is effective for tight-fitting clothing, it struggles to accurately model loose-fitting clothing that often moves far from the body. This was addressed in~\cite{Li24a} by starting from the so-called Implicit Sewing Patterns (ISP) model \cite{Li23a} that represents garments in terms of a set of individual 2D panels and 3D surfaces associated to these panels, and then applying a deformation model to the 3D surfaces so that they can deviate substantially from the body shape. These deformations are conditioned on normals estimated from an input image of the target garment, which are learned from synthetic mesh data featuring loose clothing.

The approach of~\cite{Li24a} is effective but tends to over-smooth the results. This is in part because different 3D shapes can give rise to very similar images, making it difficult to properly train a network to predict high-fidelity surface details and complex deformations from a single image. Furthermore, some parts of the garments are systematically occluded in images of people wearing them. To overcome these limitations, we introduce three diffusion schemes: 
\begin{enumerate}
 \item to learn a shape prior that captures complex garment shapes,

 \item to complement image information in occluded parts of the garments, 
 
 \item to map the 2D image to 3D and UV spaces so as to recover plausible 3D shapes by fitting them to the shape prior.

\end{enumerate}

Fig.~\ref{fig:pipe} depicts the resulting processing pipeline. We demonstrate that our method can recover realistic 3D models for various garments. It recovers more details and achieves higher reconstruction accuracy than existing approaches. Furthermore, our reconstructed meshes are readily usable by downstream applications, such as garment retargeting and texture editing. Codes are at \href{https://github.com/liren2515/DMap.git}{Github}.

\section{Related Work}
\label{sec:related}

\parag{Tight-Fitting Clothing.}
Recent advances in clothed human reconstruction have primarily focused on clothing that adheres relatively closely to the body, thereby significantly limiting the diversity of garment shapes that can be accurately represented. These approaches can be broadly categorized into two main groups.

The first category includes methods that represent the body and garment using a single 3D model. For instance, \cite{Jackson18,Zheng19} employ voxel-based representations generated by volumetric regression networks to model 3D clothed humans. Other works~\cite{Saito19a,Saito20a,He20d,Alldieck22} utilize pixel-aligned implicit functions to define 3D occupancy fields or signed distance fields for clothed humans. In~\cite{Alldieck19b,Alldieck19a}, displacement vectors or UV maps are used to capture deviations from the SMPL parametric body model~\cite{Loper15}. Similarly, approaches such as \cite{Huang20c,He21,Xiu22,Zheng21} combine parametric body models with implicit representations to enhance robustness to changes in body pose. While effective, these methods have significant limitations: because the body and clothing are jointly modeled, they cannot disentangle the garment surface from the body surface. This limitation hinders downstream applications and makes it difficult to model loose garments whose motion can behave independently of the body.

The second category consists of methods that explicitly model garments as separate surfaces that interact with the body. DeepGarment \cite{Danerek17}, MGN \cite{Bhatnagar19}, and BCNet \cite{Jiang20d} employ neural networks trained on synthetic RGB images to predict vertex positions for predefined mesh templates. Other approaches, such as \cite{Casado22,Liu23b}, optimize vertex positions based on estimated surface normals to recover fine wrinkle details. However, the reliance on predefined mesh templates inherently limits the range of garment shapes these methods can handle. Additionally, being trained on synthetic RGB data often results in poor reconstructions when applied to real-world images. To address these shortcomings, methods like SMPLicit \cite{Corona21}, DIG \cite{Li22c}, and ClothWild \cite{Moon22} leverage Signed Distance Functions (SDF) to reconstruct a wide variety of garment meshes using segmentation masks derived from RGB images. However, representing non-watertight garment surfaces with an SDF requires enclosing them within watertight surfaces of a minimum thickness, which compromises modeling accuracy and hinders subsequent refinement. To overcome this, approaches such as \cite{Guillard22b,DeLuigi23} adopt Unsigned Distance Functions (UDF). While UDF avoid watertight constraints, they introduce robustness issues: learning a sharp and clean 0-isosurface for UDF is challenging for neural networks, often resulting in inaccuracies that manifest as holes and artifacts in reconstructed models.
The Implicit Sewing Patterns (ISP) model of~\cite{Li23a} effectively addresses the issues of generality, accuracy, and robustness by its 2D pattern and UV parameterization. However, since \cite{Li23a} is trained specifically for draping, it struggles to capture the large deformations caused by dynamic garment motion.

\parag{Loose-Fitting Clothing.}
Loose-fitting clothing is significantly more challenging to reconstruct due to its large shape variations and free-flowing nature, which keeps it far from the body. Some recent works~\cite{Yang18f,Zhu20,Zhu22} rely on complex physics simulations or feature line estimation to align surface reconstructions with input images. However, their reliance on garment templates limits their generality, similar to the limitations faced by other template-based methods discussed earlier. Point-based approaches, such as those proposed in \cite{Zakharkin21,Srivastava22,Ma22a}, aim to reconstruct generic clothing. Unfortunately, point clouds are not inherently well-suited for downstream applications like cloth simulation and animation. While \cite{Srivastava22} employs a modified Poisson Surface Reconstruction (PSR) technique to generate garment surfaces from point clouds, it often produces results with incorrect geometry. ECON~\cite{Xiu23}, leveraging techniques like normal integration and shape completion, achieves visually appealing reconstructions of individuals wearing loose clothing. However, it still generates a single watertight mesh that tightly binds the body and garment together, which limits its applicability for tasks such as cloth simulation and recreation.

Recently, GaRec~\cite{Li24a} introduced a method that combines the ISP model~\cite{Li23a} with an image-conditioned deformation model to reconstruct loose-fitting clothing. Because different 3D shapes can result in similar images, it struggles to capture high-fidelity surface details and complex deformations, particularly for unseen or occluded parts that cannot be observed from monocular images. However, our method uses diffusion models to supplement, lift up, and map the 2D image observations to 3D, and fit a deformation prior to it, resulting in high-fidelity 3D reconstruction for the garment.
GarVerseLOD~\cite{Luo24} addresses garment reconstruction by creating a large-scale hierarchical dataset and training implicit models for garment recovery. However, constructing such a dataset requires significant manual effort from professional artists. In contrast, our method only relies on synthetic data generated by using readily accessible cloth simulation tools~\cite{Blender,MarvelousDesigner}, greatly reducing the need for manual intervention.

\parag{Diffusion Model.}
Diffusion models \cite{Ho20,Song21c} are a class of generative models that excel at learning complex data distributions through score matching. These models generate high-quality samples via an iterative denoising process and have demonstrated state-of-the-art performance across a variety of image-based generative tasks \cite{Dhariwal21,Chung22a,Chung22b,Rombach22}. Beyond 2D tasks, diffusion models have also been applied to various 3D domains, including text-to-3D generation \cite{Xu23c,Poole22}, image-to-3D generation \cite{Muller23,Xu23c,Anciukevivcius23,Liu23f}, and point cloud synthesis \cite{Tyszkiewic23,Melas23}. Recently, diffusion-based shape priors have been introduced for garment reconstruction~\cite{Guo24,Li24b}, which leverage UV maps for garment parameterization. However, these methods require point clouds as 3D measurements of garments and do not account for body-garment interaction during reconstruction. In contrast, our proposed method reconstructs 3D garments directly from monocular 2D images while accurately modeling both the garment and the body.


\section{Garment Representation Model}
\label{sec:disp}

In this section, we define our garment representation model, DISP. It defines a reconstruction prior we use when fitting it to images,  as discussed in the following section. DISP relies on Implicit Sewing Pattern (ISP)~\cite{Li23a} to model the garment rest geometry. ISP uses UV positional maps to model the geometry of different garments but is limited by only producing a single UV map for a specific garment, which is not enough to represent the many possible answers. To address this issue, we extend ISP into DISP by incorporating a diffusion model to capture the complex garment shapes caused by body motion. It is depicted by the gray network in Fig.~\ref{fig:pipe}. In the remainder of this section, we first describe ISP and then our proposed extension.

\subsection{Implicit Sewing Patterns}
\label{sec:isp}

\parag{Formalization.} 

Implicit Sewing Patterns (ISP)~\cite{Li23a} is a garment model based on the sewing patterns used in the fashion industry to design and manufacture clothes. A sewing pattern is made of several 2D panels along with stitch information for assembling them together. The 2D panels and the stitching are implicitly modeled using a 2D signed distance field (SDF) and a 2D label field, respectively.  For a specific garment, its corresponding latent code $\bz$, and a point $\bu$ in the 2D UV space $\Omega=[-1,1]^2$, the ISP model outputs the signed distance $s$ to the panel boundary and a label $l$ using a fully connected network $\mathcal{I}_{\Teta}$ as
\begin{equation} \label{eq:pattern}
    (s,l) = \mathcal{I}_{\Teta}(\bu,\bz) \; . 
\end{equation}
The zero crossing of the SDF defines the shape of the panel, with $s<0$ indicating that $\bu$ is within the panel and $s>0$ indicating that $\bu$ is outside the panel. The label $l$ encodes the stitch information, instructing which panel boundaries should be stitched together. To map the 2D sewing patterns to 3D surfaces, a UV parameterization function $\mathcal{A}_{\Phi}$ is learned to perform the 2D-to-3D mapping
\begin{equation} \label{eq:uv}
    \bX = \mathcal{A}_{\Phi}(\bu,\bz) \; ,
\end{equation}
where $\bX\in\mathbb{R}^3$ represents the 3D position of $\bu$. In essence, ISP registers different garments onto a unified UV space and establishes the mapping functions between points in UV space and the 3D garment surfaces. The shape of SDF's 0-crossing defines the geometry of garment in its rest state. As ISP is a differentiable representation, we can easily fit a latent code $\bz$ to arbitrary masks or contours of the panels to recover the corresponding garment geometry.

\parag{Training.}
Training ISP requires the 2D sewing patterns of rest-state 3D garments. However, they are not available in most garment datasets, e.g. CLOTH3D \cite{Bertiche20}. Following the garment flattening strategy of \cite{Li24a,Pietroni22}, we cut the garment mesh of CLOTH3D into front and back surfaces according to predefined cutting rules and then flatten them into 2D panels by minimizing an as-rigid-as-possible energy~\cite{Liu08c}.
For each garment in the dataset, a front and a back panel are generated as its sewing pattern. By using the paired 2D sewing patterns and their 3D meshes, we learn the weights of the ISP model $(\mathcal{I}_{\Teta}, \mathcal{A}_{\Phi})$ with the training procedure of \cite{Li23a}.

\subsection{Extending ISP with a Diffusion Model}
\label{sec:deformation}

For a specific garment, the UV parameterization function of ISP only produces a single UV positional map $\mathcal{U}_r$ to model its 3D shape in the rest state
\begin{equation}\label{eq:uv_rest}
    \mathcal{U}_r[u,v] = 
    \begin{cases}
        \mathcal{A}_{\Phi}(\bu,\bz), & \mbox{if } s_{\bu} \le 0\\
        \varnothing, & \mbox{if } s_{\bu} > 0
    \end{cases}
    \; ,
\end{equation}
where $\bu=(u,v)$, $s_{\bu}$ is the SDF value of $\bu$, $[\cdot,\cdot]$ denotes the standard array addressing and $\varnothing=(-1,-1,-1)$ indicates the region outside the panel. 
When dressed on the body, the garment can have various deformations due to the motion of the underlying body, which is not able to be modeled by ISP solely. Inspired by \cite{Guo24,Li24b}, we incorporate a diffusion model into ISP to capture these possible deformations by generating plausible UV maps.

Given the deformed garments worn on the body whose rest states are modeled by ISP as Eq. \ref{eq:uv_rest}, we write the corresponding UV maps
\begin{equation}\label{eq:uv_map}
    \mathcal{U}[u,v] = 
    \begin{cases}
        \bV, & \mbox{if } s_{\bu} \le 0\\
        \varnothing, & \mbox{if } s_{\bu} > 0
    \end{cases}
    \; ,
\end{equation}
where $\bV\in \mathbb{R}^3$ is the corresponding position on the deformed mesh surface for the UV point $\bu=(u,v)$. Each $\mathcal{U}$ represents a specific deformed shape for a particular garment. We use a diffusion model~\cite{Ho20} $\boldsymbol{\epsilon}_{\theta}$ to learn the distribution of plausible deformations represented by $\mathcal{U}$. 

For each garment sample, we generate its UV map $\mathcal{U}$ according to Eq. \ref{eq:uv_map}, along with a panel mask $\mathcal{M}$ as
\begin{equation}\label{eq:uv_mask}
    \mathcal{M}[u,v] = 
    \begin{cases}
        1, & \mbox{if } s_{\bu} \le 0\\
        0, & \mbox{if } s_{\bu} > 0
    \end{cases}
    \; .
\end{equation}
$\mathcal{M}$ depicts the panel shape, which itself encodes the 3D geometry of the canonical rest garment. We concatenate $\mathcal{U}$ and $\mathcal{M}$ along the channel dimension to form the training samples and train the network $\boldsymbol{\epsilon}_{\theta}$ on them.
After training, the diffusion model and ISP together form the garment model DISP.

\section{Reconstructing Garments from Monocular Images}
\label{sec:recon}

Monocular images yield 2D observations for non-occluded regions. We rely on a generative diffusion model to complement image information in occluded parts. By mapping the 2D image information to 3D and UV spaces, we then enable their fitting to DISP for the recovery of realistic 3D garments, even in parts that are not visible.

\subsection{Observations from Images}
\label{sec:observations}
Recent advances in image segmentation~\cite{Kirillov23,Ravi24}, normal estimation~\cite{Bae24,Khirodkar25} and human mesh recovery~\cite{Goel23a,Stathopoulos24}  can be used to extract accurate observations from an image of a clothed person. In this manner, 
we first segment the target garment using \cite{Kirillov23,Li20k} and estimate its normals $\bN_F$ using \cite{Khirodkar25}. To model the body underneath, we use SMPL \cite{Loper15},  which relies on two sets of parameters $(\beta,\theta)$ to describe the body shape and pose respectively. The SMPL parameters are estimated from the image by \cite{Goel23a} to infer the 3D body shape, which are then used to render front and back body part segmentations $\bS_F$ and $\bS_B$, along with front and back depth images $\bD_F^b$ and $\bD_B^b$, as shown in the top left of Fig. \ref{fig:pipe}.

To estimate the invisible normals, typically in the back, $\bN_B$ as shown in Fig. \ref{fig:pipe},  we use the estimated normal $\bN_F$ to guide a conditional diffusion model $\boldsymbol{\epsilon}_{\theta}^n$. The denoising process of $\boldsymbol{\epsilon}_{\theta}^n$ is conditioned on the visible normals $\bN_F$, the front and back segmentation images $\bS = (\bS_F, \bS_B)$, the body depth maps $\bD = (\bD_F^b, \bD_B^b)$. It is learned by minimizing the loss
\begin{equation}
    \mathcal{L} = \mathbb{E}_{t,\bN_B,\boldsymbol{\epsilon}}\| \boldsymbol{\epsilon} - \boldsymbol{\epsilon}_{\theta}^n\left( \sqrt{\bar{\alpha}_{t}}\bN_B+\sqrt{1-\bar{\alpha}_{t}}\boldsymbol{\epsilon}, \bN_F, \bS, \bD, t \right) \|^2 \; .
\end{equation}
The conditioning images of the garment and body provide information to generate plausible normals for the back of the garment. As will be shown in our experiments, the back normal estimation $\bN_B$ provides additional constraints to regularize the garment fitting process, which improves the fidelity of the reconstruction.

\subsection{Mapping from Pixel Space to UV Space and 3D Space}
\label{sec:mapping}

\begin{figure}[ht!]
    \centering
    \includegraphics[width=.47\textwidth]{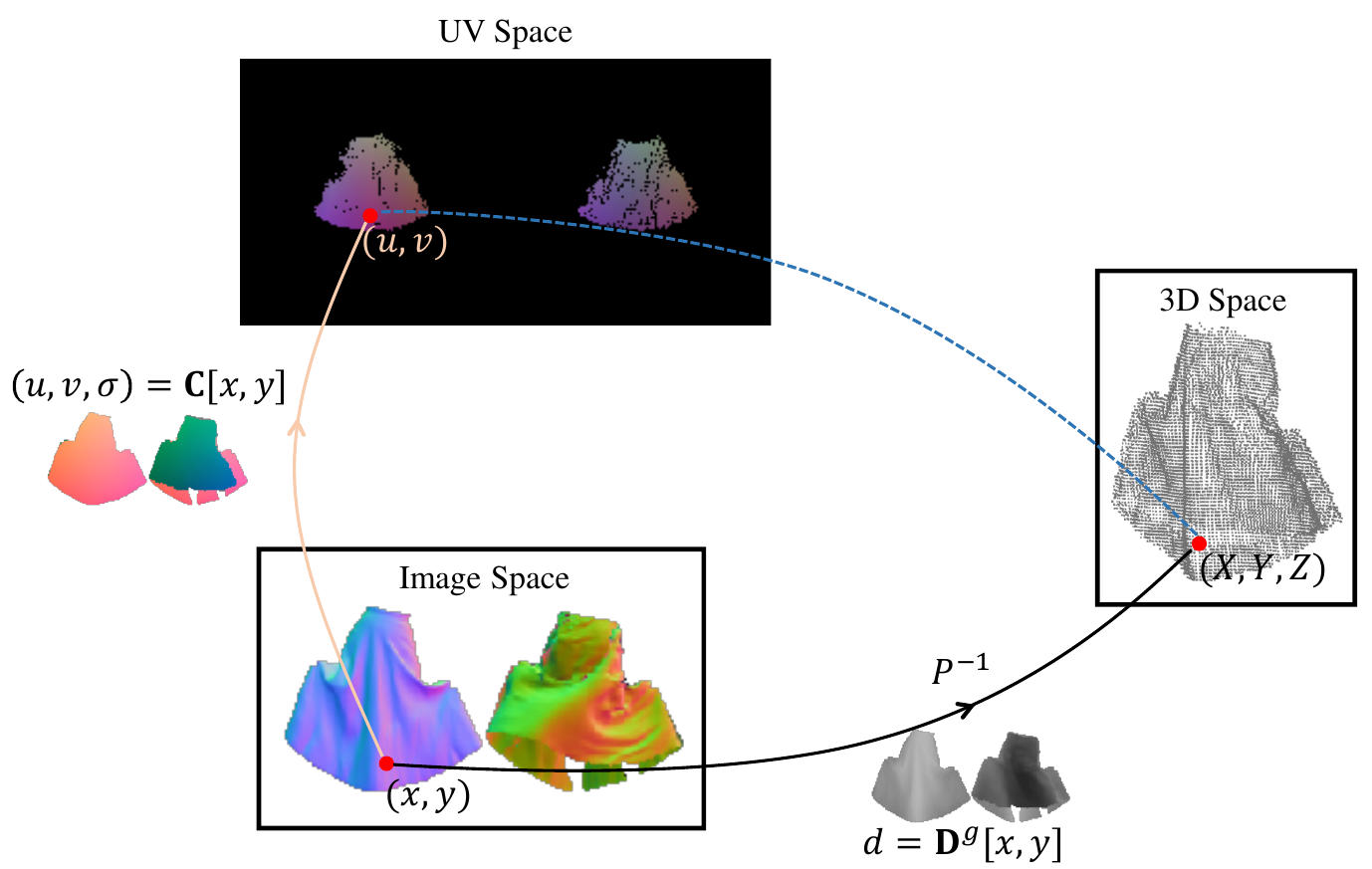}
    \caption{\textbf{Mapping between pixel, 3D, and UV spaces.} The pixel $(x,y)$ is mapped to $(X,Y,Z)$ in the 3D space using the estimated depth $d$ and the camera backprojection $P^{-1}$, and to $(u,v)$ in the UV space using the estimated UV coordinates $(u,v,\sigma)$. The dash line indicates that $(X,Y,Z)$ and $(u,v)$ are connected indirectly through $(x,y)$.}
    \label{fig:mapping}
\end{figure}

The normal estimations provide observations in pixel space, while the garment model DISP is learned in the UV space of the garment panels, and the garment surface resides in the 3D space. To reconstruct 3D garments using DISP, it is thus necessary to connect these three different spaces. To this end, we introduce a mapping function that translates image observations from the pixel space to both the UV space and the 3D space, as illustrated by Fig. \ref{fig:mapping}.

\parag{To 3D Space.} 

Since the depth and surface normal are closely related in terms of 3D geometry, we estimate the garment depth image $\bD^g$ from normal estimations $\bN$ conditioned on the body depth $\bD^b$. For the foreground pixel $(x,y)$, its absolute depth value is $d=\bD^g[x,y]$. By leveraging the camera projection $P$, we can have the 3D coordinate $(X,Y,Z)$ for each pixel
\begin{equation}\label{eq:bproj}
    (X,Y,Z) = P^{-1}(x,y,d)\; ,
\end{equation}
where $P^{-1}$ denotes the camera backprojection. Through Eq. \ref{eq:bproj}, we establish the mapping from the pixel space to the 3D space.

\parag{To UV Space.} 
Given the normal estimation $\bN$, we train a network $\boldsymbol{\epsilon}_{\theta}^m$ to predict a UV-coordinate image $\bC$ conditioned on the body part segmentation $\bS$. The pixel value of $\bC$ is 
\begin{equation}\label{eq:uv}
    \bC[x,y]  = (u,v,\sigma) \; , 
\end{equation}
where $(u,v)$ is the predicted coordinate on the UV space of the panel for pixel $(x,y)$, $\sigma$ indicates whether it belongs to the front ($\sigma > 0$) or the back ($\sigma < 0$) panel. Through Eq. \ref{eq:uv}, we establish the mapping from the pixel space to the UV space.

By assembling the results of UV and 3D mapping of Eq. \ref{eq:uv} and Eq. \ref{eq:bproj}, we can get a UV map $\Tilde{\mathcal{U}}$, where
\begin{equation}
    \Tilde{\mathcal{U}}[u,v] = P^{-1}(x,y,d) = (X,Y,Z)\; .
\end{equation}
For the positions on $\Tilde{\mathcal{U}}$ without projected points, we simply set their values to $\varnothing$. We also compute a mask $\Tilde{\mathcal{M}}$ with
$\Tilde{\mathcal{M}}[u,v] = 1$ at where a pixel is projected, and $\Tilde{\mathcal{M}}[u,v] = 0$ otherwise. Due to occlusions, both $\Tilde{\mathcal{U}}$ and $\Tilde{\mathcal{M}}$ are incomplete. In the next section, we will complete them by fitting to the priors encoded in DISP.

\parag{Training.} 
We learn the mapping function in an image-to-image translation fashion with a conditional diffusion model $\boldsymbol{\epsilon}_{\theta}^m$. For the normal estimation $\bN_F$ and $\bN_B$, $\boldsymbol{\epsilon}_{\theta}^m$ is trained to predict their UV-coordinate image $\bC_F$ and $\bC_B$, and depth images $\bD_F^g$ and $\bD_B^g$ jointly. 
The denoising process of $\boldsymbol{\epsilon}_{\theta}^m$ is conditioned on the estimated normals of the front and the back $\bN = (\bN_F, \bN_B)$, the segmentation images $\bS = (\bS_F, \bS_B)$, the body depth maps $\bD = (\bD_F^b, \bD_B^b)$, and is learned by minimizing the loss
\begin{equation}
    \mathcal{L} = \mathbb{E}_{t,\bm_0,\boldsymbol{\epsilon}}\| \boldsymbol{\epsilon} - \boldsymbol{\epsilon}_{\theta}^m\left( \sqrt{\bar{\alpha}_{t}}\bm_0+\sqrt{1-\bar{\alpha}_{t}}\boldsymbol{\epsilon}, \bN, \bS, \bD, t \right) \|^2 \; ,
\end{equation}
where $\bm_0=[\bC_F, \bC_B, \bD_F^g, \bD_B^g]$. After the training, we assemble the results for both the front and the back to produce the UV map $\Tilde{\mathcal{U}}$ and the mask $\Tilde{\mathcal{M}}$. Compared with only using the front result, this provides more observations and constraints for the fitting, resulting in a reconstruction with higher quality for both the visible and invisible parts.

\subsection{Fitting}
\label{sec:fitting}
The incomplete panel mask $\Tilde{\mathcal{M}}$ and the incomplete UV map $\Tilde{\mathcal{U}}$ provides partial information of the garment geometry and deformation, respectively. To recover a complete garment from them, we leverage the prior of DISP. We first recover the complete panel mask to recover the garment geometry in its rest shape, and then recover the complete UV map for the deformation. To remedy the synthetic-to-real domain gap and further improve the reconstruction accuracy, we rely on a post-optimization step to align the garment with image observations. 


\begin{figure}[ht!]
    \centering
    \includegraphics[width=.47\textwidth]{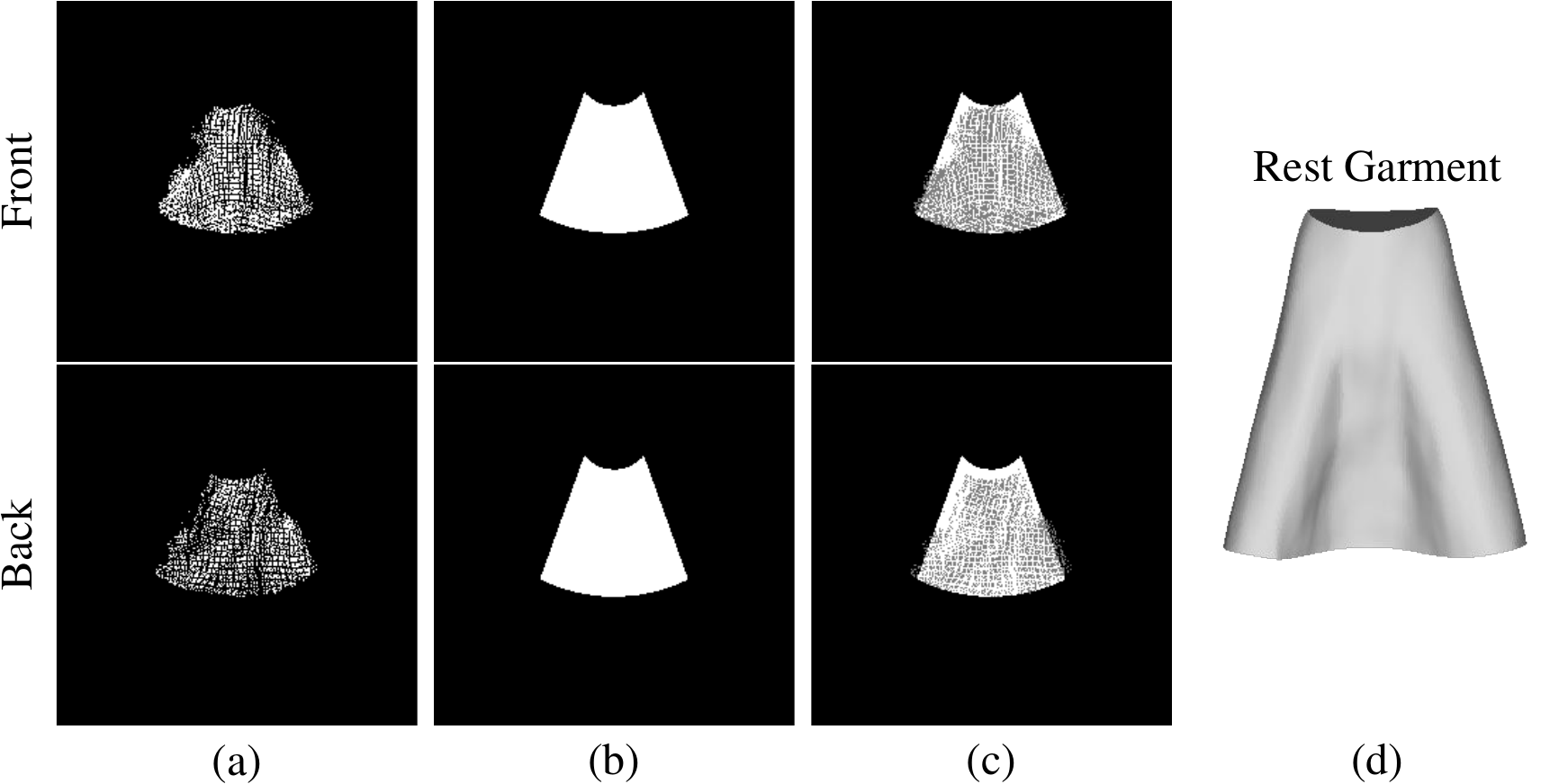}
    \caption{\textbf{Recovering garment rest geometry}. Given (a) the incomplete panel mask $\Tilde{\mathcal{M}}$, we fit (b) the complete panel mask $\mathcal{M}$ by Eq. \ref{eq:z}. (c) shows the overlay of $\Tilde{\mathcal{M}}$ in gray and $\mathcal{M}$ in white. (d) is the corresponding rest-state garment mesh $\bar{\bG}$ for (b).}
    \label{fig:mask}
    \vspace{-0.45cm}
\end{figure} 

\subsubsection{Recovering the Rest Geometry.} 
\label{sec:recoverPanel}
To recover the garment rest geometry represented by the 2D panel shape from $\Tilde{\mathcal{M}}$, we optimize the latent code $\bz$ of Eq.~\ref{eq:pattern} so that its corresponding patterns match $\Tilde{\mathcal{M}}$ as well as possible. The optimization objective is
\begin{equation} \label{eq:z} 
    \mathcal{L}(\bz) = \sum\limits_{\scalebox{.7}{$\bu\in\mathcal{M}_+$}}ReLU(s_\bu(\bz)) - \lambda_{area}\sum\limits_{\scalebox{.7}{$\bu\in\Omega$}}s_\bu(\bz) + \lambda_\bz||\bz||_2\; ,
\end{equation}
where $\mathcal{M}_+=\{\bu|\tilde{\mathcal{M}}_\bu=1, \bu\in\Omega\}$, $s_\bu(\bz)$ is the SDF value of $\bu$ computed by ISP, and $\lambda_{area}$ and $\lambda_\bz$ are the weighting constants. The first item in Eq. \ref{eq:z} ensures that the projected UV points are within the panel, while the second one penalizes large panel area to make the panel contours surround the non-zero points of $\tilde{\mathcal{M}}$ as closely as possible. This optimization produces an optimal latent code $\bz^*$ that we can use to infer a complete panel mask $\mathcal{M}$ and a rest-state garment mesh $\bar{\bG}$ as shown in Fig. \ref{fig:mask}.

\subsubsection{Recovering Deformed Geometry.} 
\label{sec:recoverUV}
The diffusion model $\epsilon_\theta$ of DISP learns the distribution of plausible deformations represented by UV maps. To recover the full UV map $\mathcal{U}$, we use the partial UV map $\tilde{\mathcal{U}}$ and the recovered panel mask $\mathcal{M}$ as the manifold guidance~\cite{Chung22a,Chung22b} in the reverse diffusion process of $\epsilon_\theta$:
\begin{small}
\begin{align}
    \nabla_{\bx_t}\log p(\bx_t|\tilde{\mathcal{U}}, \tilde{\mathcal{M}}, \mathcal{M}) &\simeq -\frac{\epsilon_\theta(\bx_t, t)}{\sigma_t}-\rho\nabla_{\bx_t}\mathcal{L}(\hat{\bx}_0, \tilde{\mathcal{U}}, \tilde{\mathcal{M}}, \mathcal{M}) \;  \label{eq:guided_loss} \; , \\
    \hat{\bx}_0 &= \frac{1}{\sqrt{\bar{\alpha}_t}}\bx_t - \sqrt{\frac{1-\bar{\alpha}_t}{\bar{\alpha}_t}}\epsilon_\theta(\bx_t, t)\; \label{eq:loss_uv}  \; ,
\end{align}
\end{small}
\hspace{-0.2em}where $\rho$ is the guidance step size. $\mathcal{L}$ is the function that measures the difference between the generated and the given UV maps and panel masks
\begin{equation} \label{eq:guidance} 
    \mathcal{L}(\hat{\bx}_0, \tilde{\mathcal{U}}, \tilde{\mathcal{M}}, \mathcal{M}) = \lVert\tilde{\mathcal{M}}*(\hat{\mathcal{U}} - \tilde{\mathcal{U}})\rVert _2 + \lVert\hat{\mathcal{M}} - \mathcal{M}\rVert _1 \; ,
\end{equation}
where $\hat{\bx}_0=[\hat{\mathcal{U}}, \hat{\mathcal{M}}]$, $\hat{\mathcal{U}}$ and $\hat{\mathcal{M}}$ refer to the generated UV map and panel mask respectively, and $*$ denotes the element-wise multiplication. With the generated UV map $\hat{\mathcal{U}}$, we update the vertices of $\bar{\bG}$ to get the recovered mesh $\bG$ as shown in the bottom-left of Fig. \ref{fig:pipe}.

\subsubsection{Garment Refinement.} 
\label{sec:post}
As the diffusion model learns the shape distribution from the garment simulation data which is generated with limited materials, external forces, and body motions, when handling in-the-wild images that are out-of-distribution, it can produce inaccurate UV maps by Eq. \ref{eq:guided_loss} and result in inaccurate garment mesh that does not align with the images. To further improve the reconstruction accuracy, we refine the recovered mesh $\bG$ in Sec. \ref{sec:recoverUV} by optimizing its vertex positions to align it with the image observations. The loss function we use is
\begin{equation} \label{eq:postrefine} 
    \mathcal{L} = \lambda_m\mathcal{L}_{mask} + \lambda_d\mathcal{L}_{depth} + \lambda_n\mathcal{L}_{normal} +\lambda_u\mathcal{L}_{uv} + \lambda_p\mathcal{L}_{phys},
\end{equation}
where $\lambda_m$, $\lambda_d$, $\lambda_n$, $\lambda_u$ and $\lambda_p$ are the weighting scalars. $\mathcal{L}_{mask}$, $\mathcal{L}_{depth}$ and $\mathcal{L}_{normal}$ penalize the difference between the rendered front and back masks, depth and normal of the mesh $\bG$ and their corresponding estimation respectively. $\mathcal{L}_{uv}$ ensures the corresponding UV map $\hat{\mathcal{U}}$ of $\bG$ aligns with the partial UV observations $\tilde{\mathcal{U}}$ by
\begin{equation} \label{eq:luv} 
    \mathcal{L}_{uv} = \lVert\tilde{\mathcal{M}}*(\hat{\mathcal{U}} - \tilde{\mathcal{U}})\rVert _2.
\end{equation}
$\mathcal{L}_{phys}$ contains a set of physics-based mesh regularization \cite{Narain12,Santesteban22} computed by using the recovered rest-state garment $\bar{\bG}$ as the reference
\begin{equation} \label{eq:lphys} 
    \mathcal{L}_{phys} = \mathcal{L}_{strain} + \mathcal{L}_{bend} + \mathcal{L}_{gravity} + \mathcal{L}_{collision} ,
\end{equation}
where $\mathcal{L}_{strain}$ is the membrane strain energy caused by the deformation, $\mathcal{L}_{bend}$ is the bending energy resulting from the folding of adjacent faces, $\mathcal{L}_{gravity}$ is the gravitational potential energy and $\mathcal{L}_{collision}$ is the penalty for body garment collision. 

However, directly optimizing the vertex positions using Eq. \ref{eq:postrefine} will lead to a suboptimal solution, as vertices are not strongly coupled.
Inspired by \cite{Ulyanov18,Gadelha21}, we introduce a multi-layer perceptron (MLP) for the optimization to update vertices by a neural displacement field. To be specific, we initialize an MLP network $f$ with learnable parameters $\phi$. Given the vertex $V$ of mesh $\bG$ and its canonical position $\bar{V}$ on $\bar{\bG}$, the network $f_\phi$ predicts its displacement by
\begin{equation} \label{eq:nn} 
    \Delta V = f_\phi(V,\bar{V}) .
\end{equation}
We use the updated vertex position $V+\Delta V$ to compute the loss of Eq. \ref{eq:postrefine} and compute the gradient with respect to $\phi$ for its learning. Since neural networks tend to learn low-frequency functions \cite{Rahaman19}, the result after this step is a bit smooth. To further recover fine surface details, we perform an additional refinement step by directly optimizing the garment mesh vertices with Eq. \ref{eq:postrefine}.


\begin{figure}[ht!]
    \centering
    \includegraphics[width=.45\textwidth]{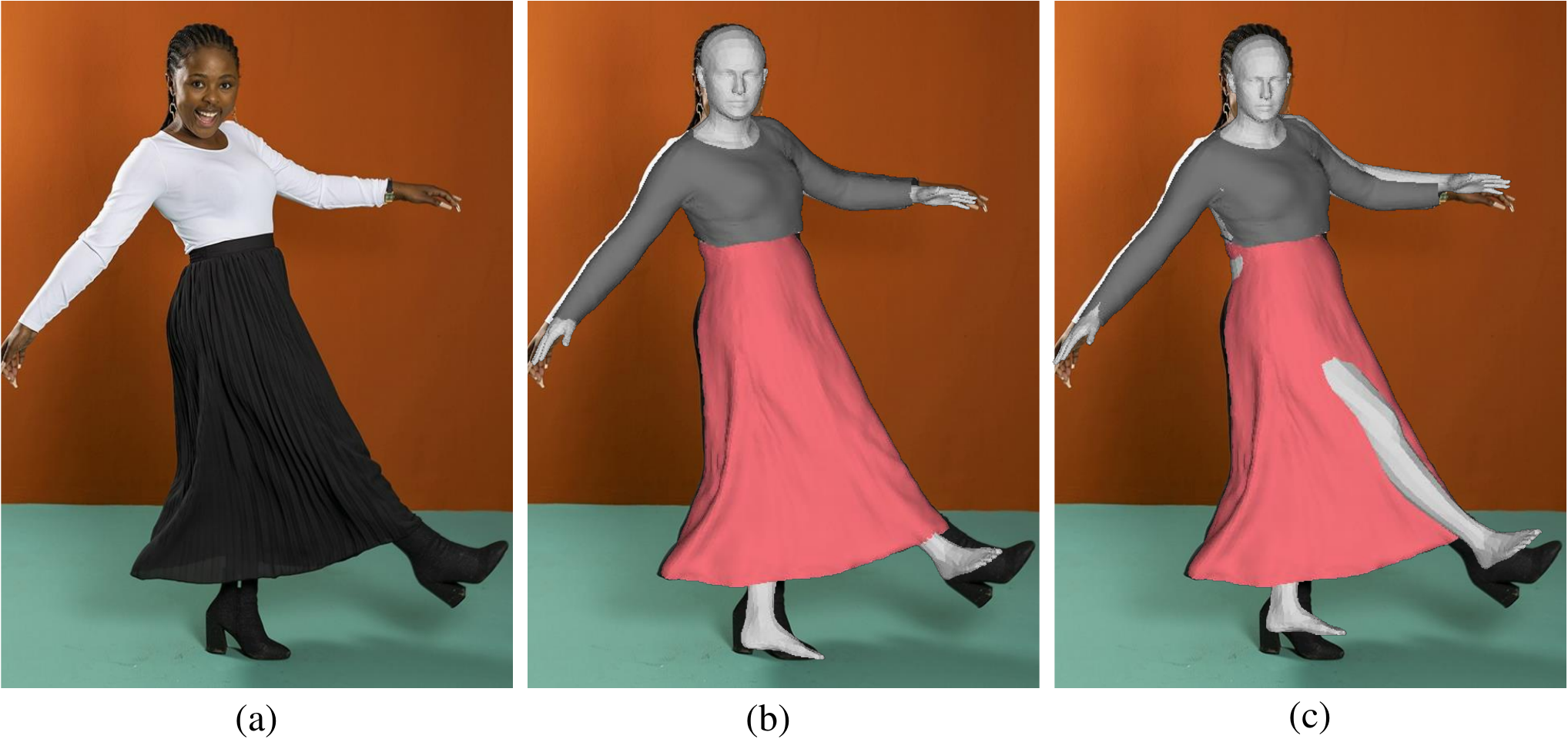}
    \caption{\textbf{Body refinement}. For the image of (a), we refine the initial body estimation of (c) by Eq. \ref{eq:bodyrefine} to improve its accuracy and align it with the image as (b).}
    \label{fig:body}
    \vspace{-0.45cm}
\end{figure} 

\subsubsection{Refining Body} 
\label{sec:postBody}
As the body estimation can be inaccurate, making it inconsistent with the garment recovery as shown in Fig.\ref{fig:body}, we refine the SMPL body parameters $\beta$ and $\theta$ by minimizing
\begin{equation} \label{eq:bodyrefine} 
    \mathcal{L}(\beta,\theta) = \lambda_m\mathcal{L}_{mask} + \lambda_o\mathcal{L}_{order} + \lambda_j\mathcal{L}_{joint} + \lambda_\beta||\beta||^2_2,
\end{equation}
where $\lambda_m$, $\lambda_o$, $\lambda_j$, and $\lambda_\beta$ are the weighting scalars. $\mathcal{L}_{mask}$ penalizes the difference between the rendered body mask with segmented mask. $\mathcal{L}_{joint}$ penalizes the difference between 2D projection of body joints and the detected 2D joints. $\mathcal{L}_{order}$ ensures that the body mesh is inside the garment mesh by enforcing the rendered depth of the body is larger than the depth of the garments $\bD_F^g$ and $\bD_B^g$ estimated in Sec. \ref{sec:mapping}
\begin{equation} \label{eq:lorder} 
    \mathcal{L}_{order} = ||ReLU(\bD_F^g - \bD_F^b + \delta)||_1 + ||ReLU(\bD_B^g - \bD_B^b + \delta)||_1 ,
\end{equation}
where $\bD_F^b$ and $\bD_B^b$ are the rendered front and back depth images of the body mesh and $\delta$ is the threshold value. Note that we first refine the body with the garment depth estimations $\bD_F^g$ and $\bD_B^g$, and then optimize the garment with the refined body as introduced in Sec. \ref{sec:post}.

\section{Experiments}

\subsection{Dataset and Evaluation Metrics}
\label{sec:data}
Due to the intricate structure of garments, collecting real 3D data with complete geometry for them is extremely difficult. Instead, we use physics-based simulation to generate garment with realistic deformations in its interaction with the underlying body.


\begin{table}[ht]
    \begin{center}
    \begin{tabular}{cccc}
     \hspace{-2mm}
      \scalebox{0.85}{
            \begin{tabular}{c | c | c | c }
            \toprule
              Skirt & CD $\downarrow$ & IoU $\uparrow$ & NC $\uparrow$ \\
             \midrule
             SMPLicit & 3.00 & 65.03 & 0.02 \\
             DrapeNet & n/a  & n/a & n/a \\
             ISP      & 2.51 & 71.10 & 0.76\\
             GaRec     & 2.00 & 93.81 & 0.80  \\
              \midrule  
             Ours & \textbf{1.21}  & \textbf{95.32} & \textbf{0.83} \\
            \bottomrule
            \end{tabular}
            }
        &
         \hspace{-4mm}
        \scalebox{0.85}{
            \begin{tabular}{c | c | c | c}
            \toprule
              Trousers & CD $\downarrow$ & IoU $\uparrow$ & NC $\uparrow$ \\
             \midrule
             SMPLicit & 1.59 & 68.19 & -0.03 \\
             DrapeNet & 1.38  & 74.23 & 0.84 \\
             ISP      & 1.53 & 57.745 & 0.85\\
             GaRec    & 1.12 & 88.52 & 0.86  \\
             \midrule
             Ours & \textbf{0.74}  & \textbf{94.00} & \textbf{0.88} \\
            \bottomrule
            \end{tabular}
            }         
        \\
        \\
         \hspace{-2mm}
        \scalebox{0.85}{
            \begin{tabular}{c | c | c | c}
            \toprule
               Shirt & CD $\downarrow$ & IoU $\uparrow$ & NC $\uparrow$ \\
             \midrule
             SMPLicit & 6.53 & 47.50 & 0.06 \\
             DrapeNet & 1.93  & 77.15 & 0.84 \\
             ISP      & 1.92 & 68.64 & 0.83\\
             GaRec     & 1.20 & 93.20 & 0.83  \\
             \midrule
             Ours & \textbf{0.85}  & \textbf{94.02} & \textbf{0.89} \\
            \bottomrule
            \end{tabular}
            }  
        &
         \hspace{-4mm}
        \scalebox{0.85}{
            \begin{tabular}{c | c | c | c}
            \toprule
              Open Shirt & CD $\downarrow$ & IoU $\uparrow$ & NC $\uparrow$ \\
             \midrule
             SMPLicit & 2.37 & 61.27 & -0.08 \\
             DrapeNet & 1.76  & 73.56 & 0.78 \\
             ISP      & 1.90 & 69.27 & 0.77\\
             GaRec     & 1.46 & \textbf{92.81} & 0.73  \\
             \midrule
             Ours & \textbf{1.19}  & 92.48 & \textbf{0.82} \\
            \bottomrule
            \end{tabular}
            }  
      \end{tabular}
      \end{center}
      \caption{\textbf{Quantitative comparisons.} Our method outperforms SMPLicit, DrapeNet, ISP and GaRec in terms of CD, IoU and NC on different garment categories. The unit of CD is cm.
      }
      \label{tab:synthetic}
      \vspace{-0.45cm}
\end{table}

CLOTH3D \cite{Bertiche20} is a synthetic dataset, with 3D garments draped on T-posed SMPL bodies \cite{Loper14}. For each clothing category, including shirt, open shirt, skirt and trousers, we randomly select 33 samples. Each pair of garment and body models is simulated with the motion data sourced from the dance category of the AMASS dataset \cite{Mahmood19}. The motion sequences are generated by using Blender \cite{Blender} and Marvelous Designer \cite{MarvelousDesigner}. Additional pins are manually set for open shirt, skirt and trousers to avoid sliding during the simulation. For each body sample in the sequence, we randomly rotate it and render its front and back body part segmentation and depth images. For the corresponding garment sample, we rotate it with the same angle and render its front and back normal and depth images. Its front and back UV coordinate images are generated using the UV parameterization of ISP. For each garment category, we randomly select 30 pairs of garment and body for training and use the rest pairs for the evaluation.

To evaluate the quality of garment reconstruction, we use the Chamfer Distance (CD) and the Normal Consistency (NC) between the ground truth and the recovered garment mesh, and the Intersection over Union (IoU) between the ground truth mask and the rendered mask of reconstructed garment mesh. The quantitative comparison is conducted on the test set of the synthetic data, and the qualitative evaluation is conducted on in-the-wild images.
 
\subsection{Results}
In Table \ref{tab:synthetic}, we present the quantitative comparison between our method and the state-of-the-art approaches: SMPLicit~\cite{Corona21}, DrapeNet \cite{DeLuigi23}, ISP~\cite{Li23a}, and GaRec~\cite{Li24a}. Our method achieves significantly better performance across all garment categories—Skirt, Trousers, Shirt, and Open Shirt—in terms of CD, IoU and NC.

\begin{figure*}[ht!]
    \centering
    \includegraphics[width=.85\textwidth]{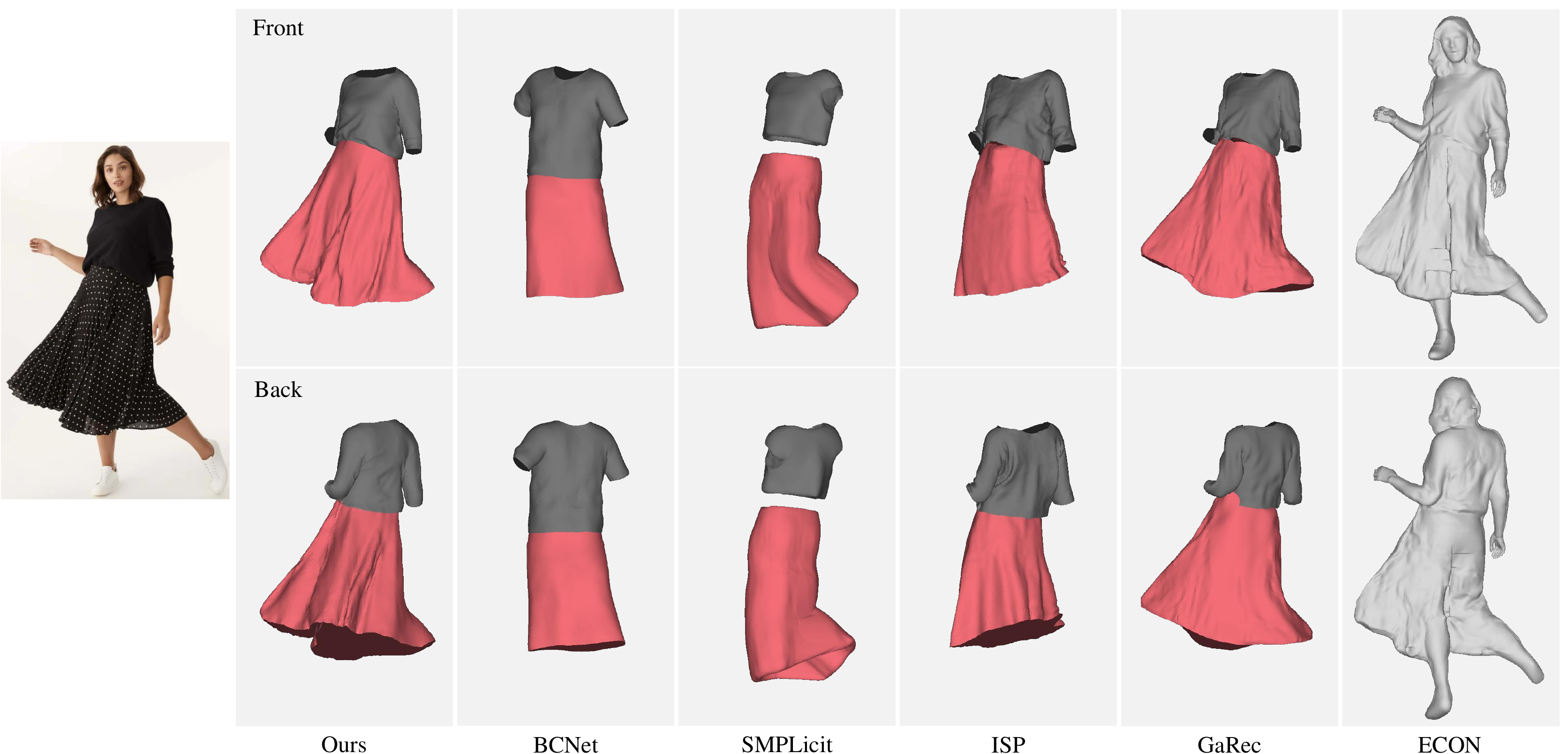}
    \vspace{-0.35cm}
    \caption{\textbf{Qualitative comparison with state-of-the-art methods.} The top and bottom rows show the front and the back of the reconstructions produced by our method, BCNet \cite{Jiang20d}, SMPLicit \cite{Corona21}, ISP \cite{Li23a}, GaRec \cite{Li24a} and ECON \cite{Xiu23}, respectively.}
    \label{fig:compare}
\end{figure*}

\begin{figure*}[ht!]
    \centering
    \includegraphics[width=.98\textwidth]{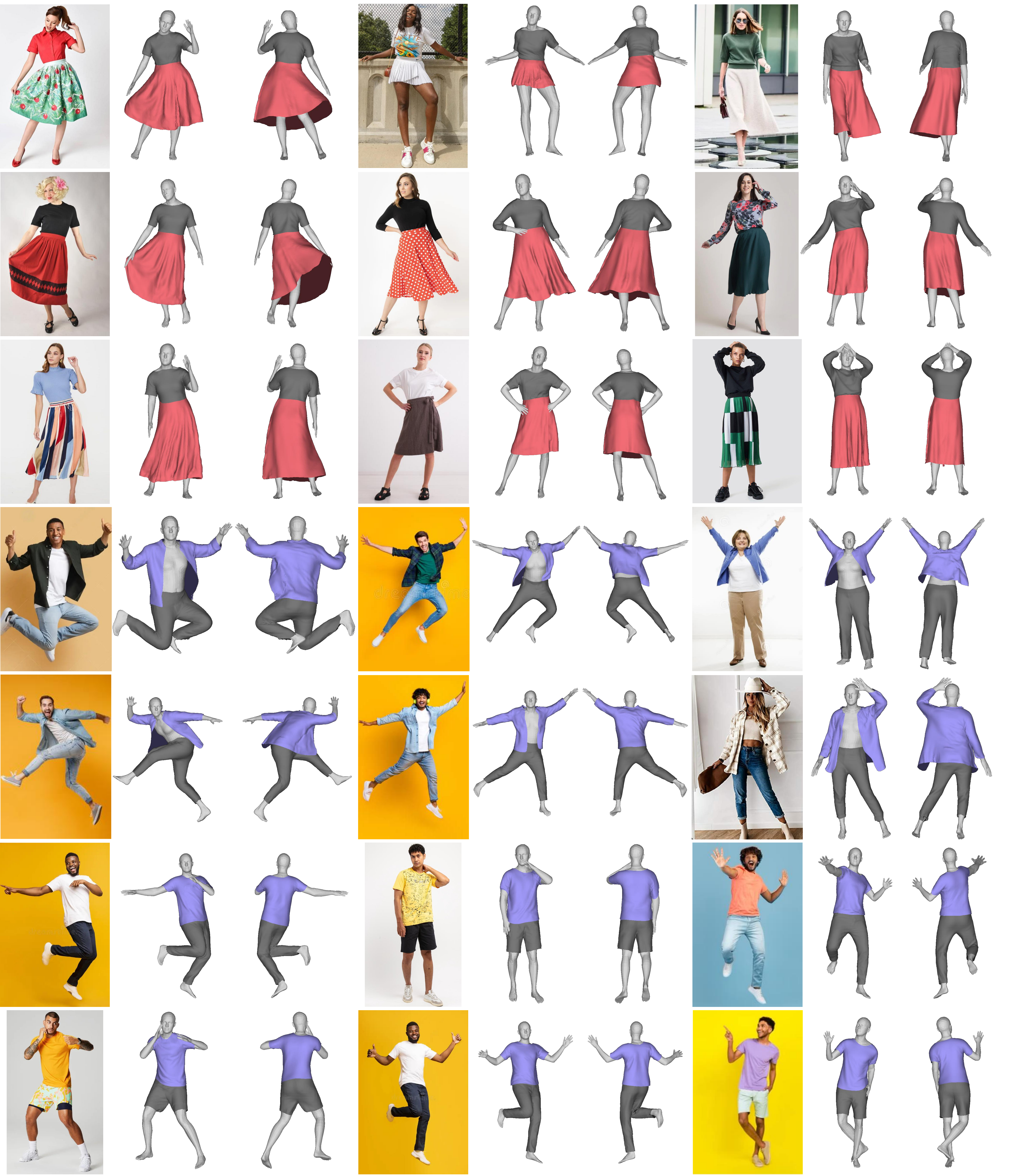}
    \caption{\textbf{More reconstruction results for in-the-wild images.} Our method can handle both the tight-fitting and the loose-fitting garments and recover high-fidelity 3D meshes for them.}
    \label{fig:supp_fitting}
    \vspace{-0.45cm}
\end{figure*}

Fig. \ref{fig:compare} provides a qualitative comparison of the results reconstructed from the in-the-wild image. Methods like BCNet \cite{Jiang20d}, SMPLicit, and ISP rely solely on the SMPL body model's skinning function to deform garments, which limits them to generating results tightly adhered to the body surface. ECON \cite{Xiu23} and GaRec, on the other hand, can recover garments that stand away from the body. However, ECON produces a single watertight mesh that models both the body and garment as a single entity. While GaRec generates standalone garment surfaces, its recovered meshes appear overly flat and lack realistic folds or creases.
In contrast, our method, leveraging the proposed fitting approach that incorporates back normal estimation and DISP priors, faithfully reconstructs garment meshes with high-fidelity wrinkle details, both on the front and back of the garment. Fig. \ref{fig:fitting} and \ref{fig:supp_fitting} provides more results of our method applied to in-the-wild images, demonstrating its ability to produce realistic 3D meshes with fine details for both tight-fitting and loose-fitting garments.

\subsection{Ablation Study}
\label{sec:ablation}
\begin{figure*}[ht!]
    \centering
    \includegraphics[width=.95\textwidth]{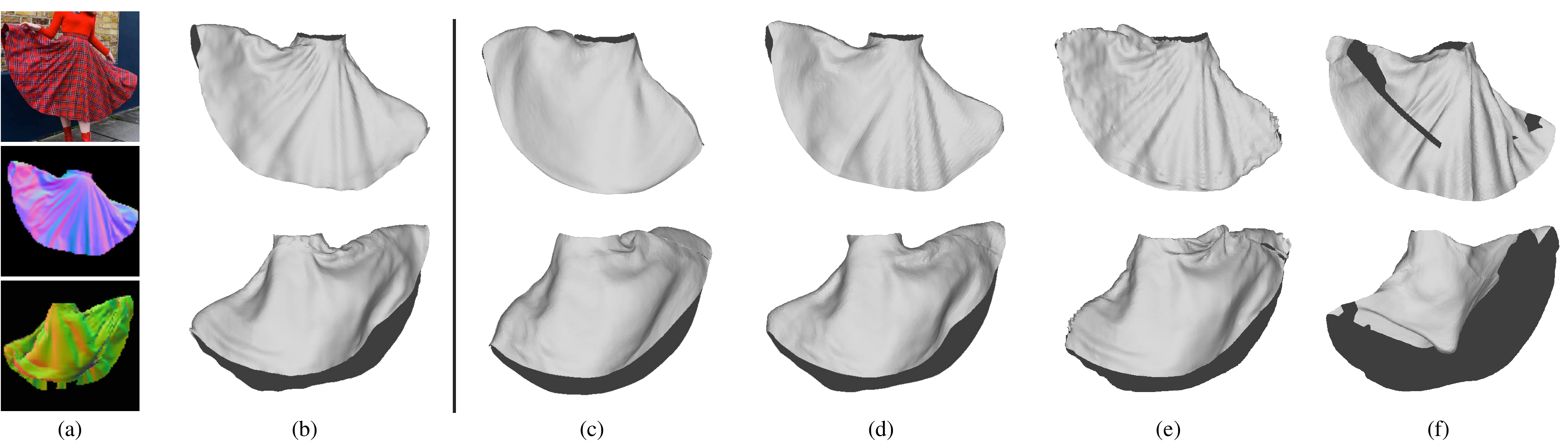}
    \vspace{-0.35cm}
    \caption{\textbf{Ablation study}. (a) The input image and its normal estimations for the front and back. (b) Our full reconstruction. (c) Reconstruction without post-refinement. (d) Reconstruction refined by optimizing only the neural displacement field. (e) Reconstruction refined by optimizing only the vertex positions. (f) Reconstruction using only the front normal.}
    \label{fig:ablation}
    \vspace{-0.25cm}
\end{figure*}

Fig. \ref{fig:ablation} presents the ablation study of our fitting method. As shown in Fig. \ref{fig:ablation}(c), the initial reconstruction, without the post-refinement step described in Section \ref{sec:post}, fails to fully align with the input image shown in Fig. \ref{fig:ablation}(a). Incorporating a neural displacement field to optimize the initial mesh improves reconstruction accuracy, as seen in Fig. \ref{fig:ablation}(d). Further refinement by directly optimizing vertex positions enhances the wrinkle details, as illustrated in Fig. \ref{fig:ablation}(b). However, applying post-refinement without first optimizing the neural displacement field (Fig. \ref{fig:ablation}(e)) struggles to recover an accurate shape, as each vertex is optimized independently, leading to suboptimal results. Finally, Fig. \ref{fig:ablation}(f) shows the outcome when only the front normal estimation is used throughout the fitting process described in Section \ref{sec:fitting}. The lack of constraints for the back surface results in unrealistic deformations on the back. The corresponding quantitative results are provided in the supplementary materials.

\subsection{Downstream Applications}

\begin{figure*}[ht]
	\centering
		\begin{minipage}{.485\textwidth}
        \centering
        \includegraphics[width=.99\textwidth]{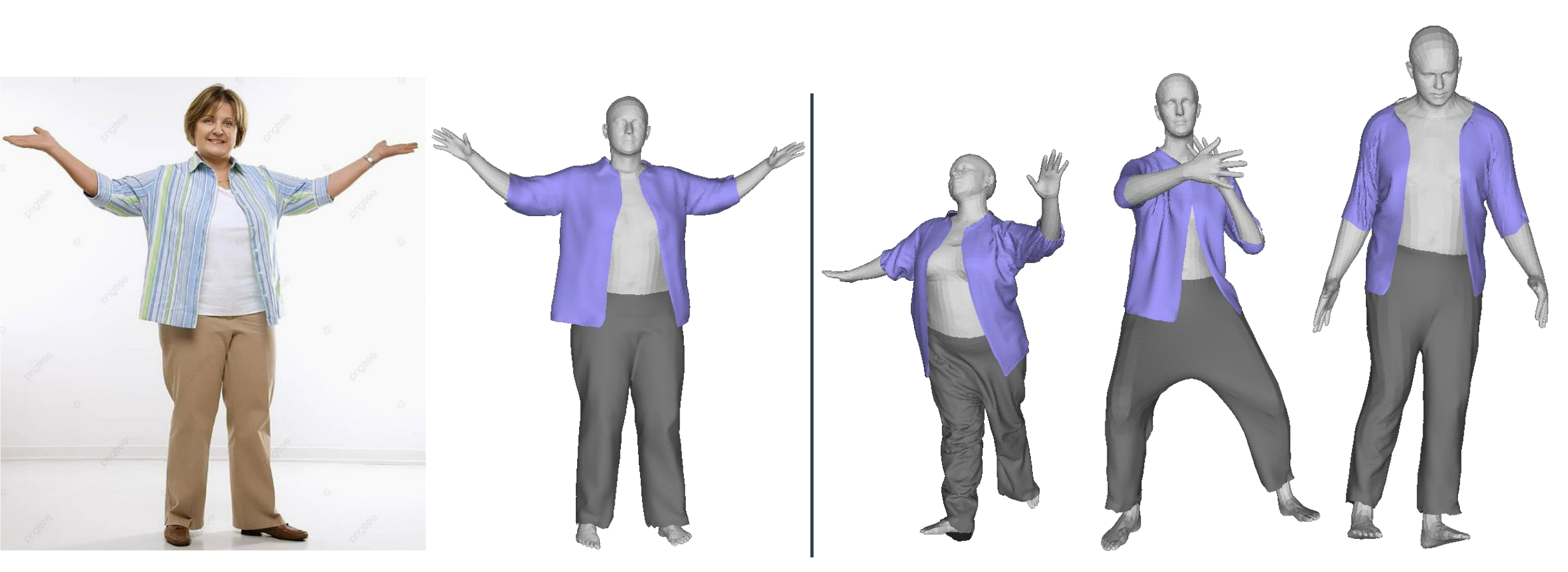}
        \caption{\textbf{Retargeting.} Left: The input image and our reconstructions. Right: The reconstructed garments are transferred to body with different poses and shapes.}
        \label{fig:repose}
    \end{minipage}
	\begin{minipage}{.485\textwidth}
		\centering
        \includegraphics[width=.99\textwidth]{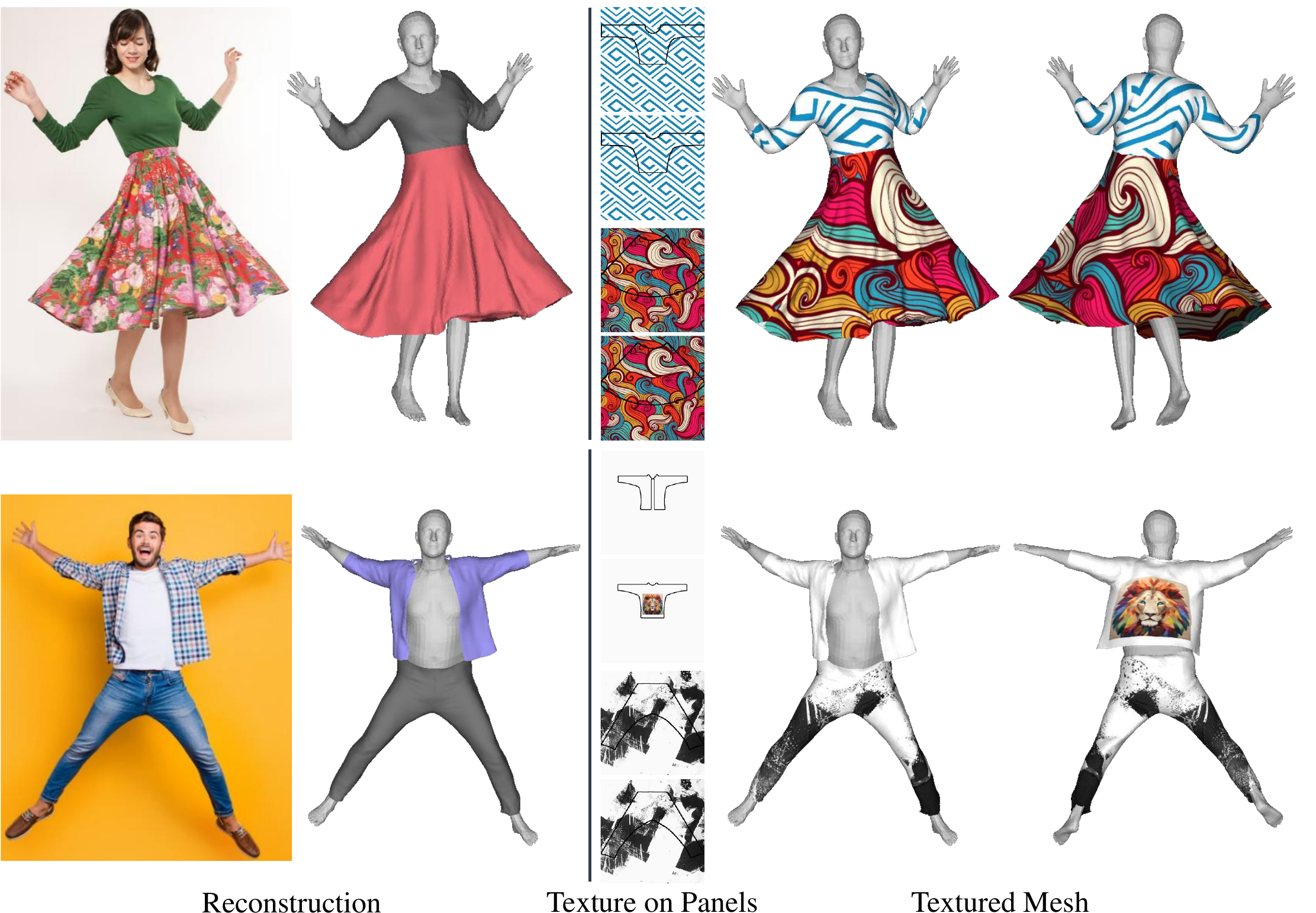}
        \vspace{-0.35cm}
        \caption{\textbf{Texture editing}. By simply drawing figures or patterns on the recovered 2D panels, we can directly edit the texture of the recovered garment mesh.}
        \label{fig:texture}
	\end{minipage}%
\end{figure*}

\parag{Retargeting} Since our method produces separate models for the garment and the underlying body, we can easily repose it on the new body. In Fig. \ref{fig:repose}, we show the retargeting results for the reconstructed open shirt and trousers by transferring them onto bodies with different poses and shapes. Accurate reconstruction of garments results in realistic retargeting. 

\parag{Texture Editing} Since we reconstruct both the 3D model and the corresponding 2D panels for garment, we can easily realize texture editing. As shown in Fig. \ref{fig:texture}, by painting patterns or drawing specific figures onto the recovered panels, the mesh will show the texture on the corresponding position.
\section{Conclusion}
We have presented a novel approach for recovering realistic 3D garment meshes from monocular images. Our method leverages Implicit Sewing Patterns (ISP) and a generative diffusion model to learn plausible garment shape priors defined in a 2D UV space. By utilizing diffusion schemes, we complement 2D observations for the occluded parts of the garments and lift them into 3D space. Additionally, we design a diffusion-based mapping across 2D, 3D, and UV space, enabling the alignment of learned priors with image observations to produce accurate 3D garment reconstructions. Our method outperforms existing approaches across different types of garments, and the resulting reconstructions are readily applicable to downstream tasks, such as garment retargeting and texture editing.

\begin{figure*}[ht!]
    \centering
    \includegraphics[width=1.\textwidth]{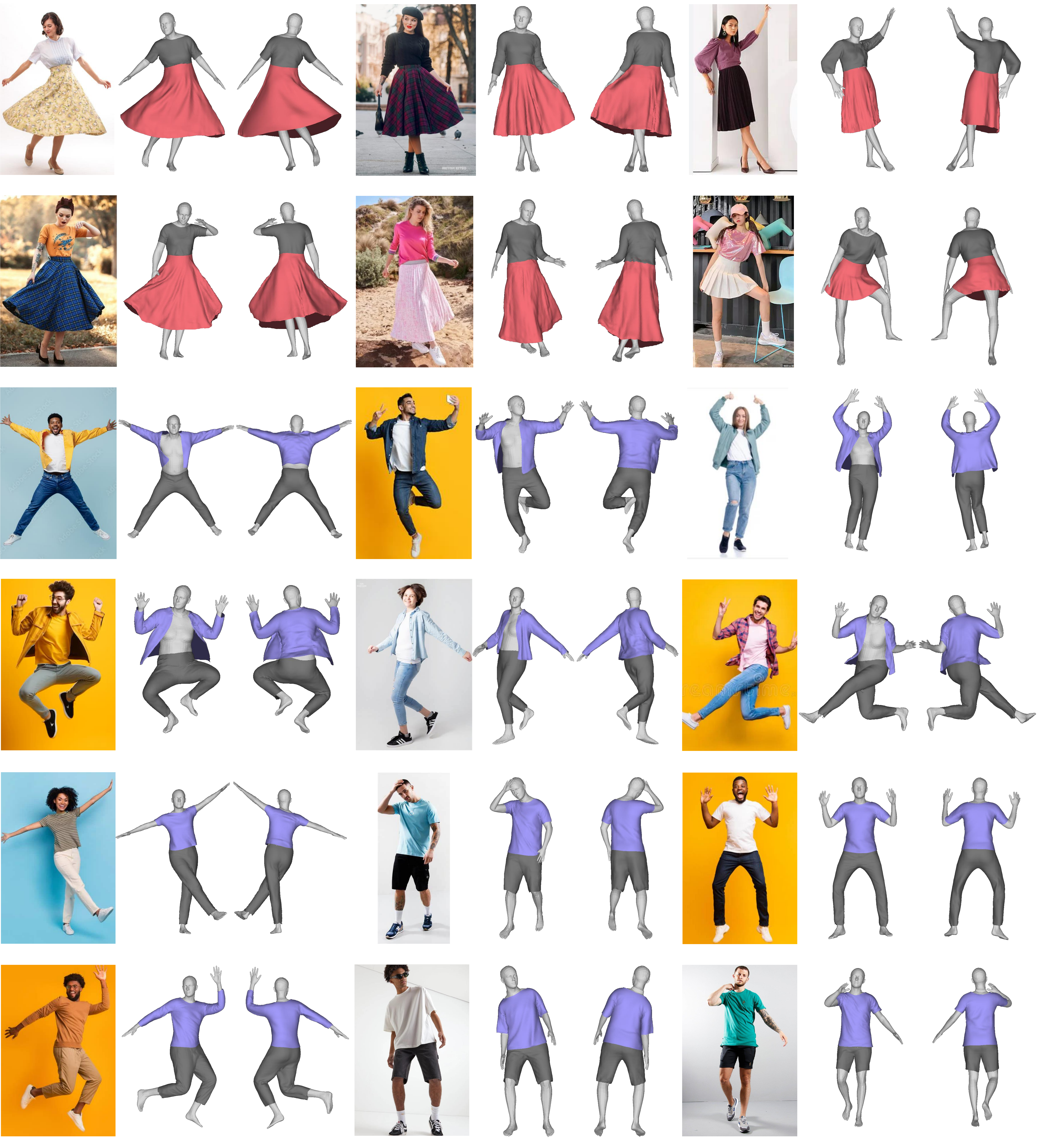}
    \caption{\textbf{Reconstruction for in-the-wild images.}  Our method can recover realistic 3D models for diverse garments in different shapes and deformations.}
    \label{fig:fitting}
\end{figure*}

\parag{Limitations} 
While our method is capable of producing realistic 3D reconstructions for a wide variety of garments, it has certain limitations. 
As shown in the middle example of the fourth row in Fig. \ref{fig:fitting}, it is challenging for our method to capture very small wrinkles. This limitation arises because our normal loss relies on a differentiable renderer~\cite{Pytorch3D}, which uses interpolation and approximations for normal and gradient computation. These approximations tend to smooth out high-frequency geometric details. Additionally, since small wrinkles contribute only marginally to the overall loss, their gradients can be overwhelmed during optimization. Together, these factors explain the observed results.
Beside, our method cannot currently handle garments with multi-layered structures, such as ruffle-layered skirts. A potential solution could involve incorporating additional panels into ISP to support layered designs. Furthermore, our approach requires full-body images of clothed individuals and, therefore, cannot handle images with partial garments or profile views. Due to the inherent ill-posed nature of 3D reconstruction from a single image, our method also cannot address depth ambiguity and does not fully capture the physical behavior of garments. The result for video input can also seem jittery. In future work, we aim to address this problem by modeling garment deformations over time using video sequences.

\section*{Acknowledgments}
This work was supported in part by the Swiss National Science Foundation grant and by the Metaverse Center Grant from the MBZUAI Research Office.

\bibliographystyle{ACM-Reference-Format}
\bibliography{string,geom,graphics,learning,vision,biomed,misc,new_ref}

\clearpage

\clearpage

\end{document}